

\magnification=\magstep0
\hsize=13.5 cm               
\vsize=19.0 cm               
\baselineskip=12 pt plus 1 pt minus 1 pt  
\parindent=0.5 cm  
\hoffset=1.3 cm      
\voffset=2.5 cm      
\font\twelvebf=cmbx10 at 12truept 
\font\twelverm=cmr10 at 12truept 
\overfullrule=0pt
\nopagenumbers    
%
\newtoks\leftheadline \leftheadline={\hfill {\eightit Authors' name}
\hfill}
\newtoks\rightheadline \rightheadline={\hfill {\eightit the running title}
 \hfill}
\newtoks\firstheadline \firstheadline={{\eightrm Bull. Astron. Soc.
India (2000) {\eightbf xx,} } \hfill}
\def\makeheadline{\vbox to 0pt{\vskip -22.5pt
\line{\vbox to 8.5 pt{}\ifnum\pageno=1\the\firstheadline\else%
\ifodd\pageno\the\rightheadline\else%
\the\leftheadline\fi\fi}\vss}\nointerlineskip}
%
\font\eightrm=cmr8  \font\eighti=cmmi8  \font\eightsy=cmsy8
\font\eightbf=cmbx8 \font\eighttt=cmtt8 \font\eightit=cmti8
\font\eightsl=cmsl8
\font\sixrm=cmr6    \font\sixi=cmmi6    \font\sixsy=cmsy6
\font\sixbf=cmbx6
%
\def\eightpoint{\def\rm{\fam0\eightrm}
\textfont0=\eightrm \scriptfont0=\sixrm \scriptscriptfont0=\fiverm
\textfont1=\eighti  \scriptfont1=\sixi  \scriptscriptfont1=\fivei
\textfont2=\eightsy \scriptfont2=\sixsy \scriptscriptfont2=\fivesy
\textfont3=\tenex   \scriptfont3=\tenex \scriptscriptfont3=\tenex
\textfont\itfam=\eightit  \def\it{\fam\itfam\eightit}%
\textfont\slfam=\eightsl  \def\sl{\fam\slfam\eightsl}%
\textfont\ttfam=\eighttt  \def\tt{\fam\ttfam\eighttt}%
\textfont\bffam=\eightbf  \scriptfont\bffam=\sixbf
\scriptscriptfont\bffam=\fivebf \def\bf{\fam\bffam\eightbf}%
\normalbaselineskip=10pt plus 0.1 pt minus 0.1 pt
\normalbaselines
\abovedisplayskip=10pt plus 2.4pt minus 7pt
\belowdisplayskip=10pt plus 2.4pt minus 7pt
\belowdisplayshortskip=5.6pt plus 2.4pt minus 3.2pt \rm}
%
%
\def\leftdisplay#1\eqno#2$${\line{\indent\indent\indent%
$\displaystyle{#1}$\hfil #2}$$}
\everydisplay{\leftdisplay}
%
\def\frac#1#2{{#1\over#2}}


%
%
\def\pmb#1{\setbox0=\hbox{$#1$}\kern-0.015em\copy0\kern-\wd0%
\kern0.03em\copy0\kern-\wd0\kern-0.015em\raise0.03em\box0}
%
\pageno=1
\vglue 60 pt  
%
\leftline{\twelvebf  Luminosity Effects in Projected Fractals}

%
\smallskip
\vskip 46 pt  
\leftline{\twelverm R. Thieberger and E.A. Spiegel} 
\vskip 4 pt
\leftline{\eightit Department of Astronomy, Columbia University, New
York, 10021, USA }
%
%
\vskip 20 pt 
%
%
\leftheadline={\hfill {\eightit R. Thieberger \& E.A. Spiegel} \hfill}
\rightheadline={\hfill {\eightit  Projection of Fractals and the
Distribution of Galaxies
}  \hfill}

%
{\parindent=0cm\leftskip=1.5 cm

{\bf Abstract.} \noindent The use of two-dimensional catalogues in
unraveling the large-scale distribution of extra-galactic objects
can reveal more information than has been supposed if the objects
have approximate scaling properties such as observations suggest. After a
brief general discussion of this issue, we turn to specific examples of
projected fractals for the case where the objects studied have a Schechter
luminosity function.  We analyze the effects of projection on the
characteristics of such a fractal distribution.  Our results indicate
that two-dimensional catalogues of sources could be of value in detecting
the effects of luminosity functions as well as of large-scale structure.

}                                 
%
%
%
\vskip 20 pt
\centerline{\bf 1. Introduction}
\bigskip
\noindent
What we now call large scale structure --- the patterns revealed
in the spatial distribution of galaxies --- was first detected
in attempts to quantify the distribution of galaxies on the sky as
revealed in position catalogues.  Those first studies were mainly
statistical and were carried out in terms of correlation
functions of the galaxy positions (Neyman, {\it et al.}, 1953; Neyman
and Scott, 1959; Peebles 1980), though one should also note the less
objective but revealing studies by de Vaucouleurs (1970). In the early
statistical work, the galaxies were considered to be points, as they were
in the related studies of catalogues of radio sources under the rubric of
$\log N - \log S$ studies (Ryle and Clarke, 1961; Peebles, 1983); that
picture is also adopted here.  From those positional studies,
Mandelbrot (1975) recognized that the galaxy distribution is well modeled
by what he has called a fractal distribution~\footnote{*}{We use the term
{\it fractal} in its most general sense to include the special cases
called multifractals and monofractals, which terms have come into
common use in recent years.}  (Mandelbrot, 1982).

In more recent times, three-dimensional
galaxy catalogues have been made available in which the three coordinates
assigned to each galaxy are two coordinates on the sky and a redshift.
Though these data are not as abundant as those in the
two-dimensional catalogues, they nevertheless support the findings of
approximate self-similarity in the large-scale distribution of galaxies
within prescribed scale intervals (Provenzale, 1991). However such
conclusions depend on the transformation from redshift space to
geometrical space, which is not completely accurate because of the
peculiar velocities of the objects in question.  Above all, since the
data in two-dimensional catalogues are more plentiful than those in three
dimensions and are likely to remain so for a time (Cress, et al., 1996),
it seems worthwhile to discuss further the ways in which the
two-dimensional data may be used to unravel the nature of the
three-dimensional distribution of extra-galactic objects.

The basic issue in analyzing a two-dimensional catalogue goes back to
work on stellar statistics (Trumpler and Weaver, 1953), with some added
complications.  In stellar statistics, the effect of fractality was not
considered much, though in the galaxy case it is important.  What
was significant in the stellar studies was the effect of the luminosity
distribution of the observed objects on the projected distributions, an
issue is our main concern here. However, rather than deal
with the inverse problem implied by the integral equation known as the
fundamental equation of stellar statistics (Trumpler and Weaver, 1953), we
shall study the effects of projection by directly simulating the
projection of some simple point sets.  In this way, we can explicitly
study the role of luminosity functions in the projection of fractals.
However, for the present, we shall leave out of account some other
features of the galaxy case that remain to be considered in detail, namely
source evolution and cosmological effects (Spedalere and Schucking, 1980;
Ribeiro, 1995).  Our main concern then is not so far from that of the
study of stellar statistics: disentangling the luminosity function of
the observed objects from their complex distribution on the sky.
Naturally, the stellar case has difficulties of its own, such as
interstellar absorption. Although an analogous issue has been examined
for the galaxy distribution by Durrer, et al. (1997), that is also
something for another exploration.

One of our motivations for raising these issues comes from our
interest in trying to distinguish the different scaling regimes
that may exist in the galaxy distribution, though on different
length scales.  It has been suggested (Murante {\it
et al.}, 1998) that three scaling regimes may be discerned in the
distribution of galaxies: on the smallest scales the results are
consistent with a distribution of density singularities; in
the intermediate range, there is scaling behaviour suggestive of
flat structures such as Zeldovitch (1970) favored; on the largest
scales, the data indicate a homogeneous distribution of galaxies
with nonfractal behavior.  This categorization is based on the
results of an analysis of three-dimensional catalogues. However,
it has to be admitted that those data may as yet not be adequate
for clearly deciding such issues and that is why we wish to
consider whether we may reliably use two-dimensional catalogues for
immediate needs.

Before describing how projection effects are to be treated, we
devote the next section to a brief outline of how structure in the
galaxy distribution may be quantified.  Then we turn to the distinction
between theoretical predictions and numerical calculations of the
effects of projection on the basis of simple examples.   To understand
this, we must try to minimize the influence of details such as the number
of points in the sets and the parameters of the luminosity functions.

Then we go on to investigate the influence of
luminosity effects by imposing a Schechter luminosity
function on the objects of study and attempt to see how
this influences the distribution of the projected objects,
or rather of their apparent magnitudes. We conclude that
luminosity effects can indeed make themselves felt in the
macroscopic parameters of the projected fractals and suggest
how these issues may be included in quantitative analyses of
two-dimensional catalogues.
\vskip 20 pt
\centerline{ {\bf 2. Quantifying Cosmic Scaling}    }
\bigskip
\noindent In a rarefied gas, the usual definition of density
breaks down as we go to the limit of small volumes.  The situation
is even worse for fractals where the density as usually understood
is singular, as we shall see presently.  Since the correlation
functions used in the early analyses of the galaxy distribution
(Neyman Scott, 1959;  Peebles 1980) are defined in terms of the
density,  this has led to questioning of the use of correlations
for the study of fractal sets (Bhavsar, 1980; Coleman and Pietronero
1992).  Fortunately, only a slight change in approach is needed to obtain
a more robust characterization of the structures of point sets by
way of the {\it correlation integral} (Grassberger and Procaccia,
1983; Paladin and Vulpiani, 1984; Borgani, 1995).  Indeed, the
change seems so slight that one may be surprised at the difference
it makes. Nevertheless, explicit calculations performed on analytically
understood point sets have shown that the correlation integral is the more
reliable tool of analysis of the properties of point distributions
(Thieberger {\it et al.} 1990).  For these reasons, the correlation
integral has been used increasingly in the study of large-scale structure
(Provenzale, 1991) and, in this section, we briefly summarize its
properties and generalizations.  A deeper discussion of this question
is provided by Bessis and Fournier (1989).

Consider then a set of $N$ galaxies --- or points --- for which the
correlation integral may be defined as $$
C_2(r)= {1\over {N(N-1)}}\sum_i \sum_{j\ne i}
\Theta(r - |{\bf X}_i - {\bf X}_j|) \eqno(2.1)$$ where
$\Theta$ is the Heaviside function and the summations are over $N-1$
galaxies with coordinates ${\bf X}_j$, $j\ne i$ (Grassberger and Procaccia
1983).  Modifications of this formula to allow for the effects of the
finiteness of the sample may be introduced when dealing with real data
(Murante {\it et al.}, 1997; Kerscher, 1999), but we pass over such
technical details here.

We may interpret $C_2(r)$ as ${\cal N}(r)/N$ where ${\cal N}(r)$ is the
average number of galaxies within a distance $r$ of a typical galaxy in
the set.  As $r$ goes to zero, $C_2$ should go to zero and, for general
distributions, we express this conclusion as $C_2 \propto r^{D_2}$.
The exponent $D_2$ is called the correlation dimension
and it is necessarily $\le 3$ for an embedding space of dimension three.
When $D_2$ is not an integer, the distribution is called fractal
(Mandelbrot, 1982) and, in that case, the density, $n(r) \propto {\cal
N}/R^3 \propto r^{D_2-3}$, has a singularity, as mentioned above.

As with correlation functions, it is possible to study higher
order statistics.  To generalize the correlation integral
formalism, one introduces (Paladin and Vulpiani 1987) $$ C_q(r)=
\left({1\over {N}}\sum_i \left[ {1\over {N-1}} \sum_{j\ne i}
\Theta(r - |{\bf X}_i - {\bf X}_j|)\right]^{q-1}\right)^{1\over
q-1} \ , \eqno(2.2)$$ where $q$ is a parameter that defines the
order of the moment. For $q=2$, a two-point probability
(second-order moment) is evaluated and the standard correlation
integral is recovered.  For any integer value of $q$, $C_q(r)$ is the
fraction of $q$-tuples in the set whose members lie within a
distance $r$ of one another.  For sufficiently small $r$, $C_q$
will go to zero for $q > 1$ and, for a typically well-behaved set,
it will vanish like $r^{D_q}$, where the index $D_q$ is called a
generalized or Renyi dimension (Renyi 1970, Halsey {\it et al.}
1986).

A fractal whose dimensions are all the same ($D_q$ independent of $q$) is
called a homogeneous fractal, or monofractal.  The more general cases
with $D_q$ depending on $q$ are called multifractals.  Some authors
reserve the term fractal for the special case of a monofractal but, as
mentioned at the outset, we here retain the general sense of the term,
with the multifractal as a particular case.  For small $r$, in leading
order, a plot of $\log C_q$ versus $\log r$ is approximately linear and
the slope is $D_q$.  The dimensions are therefore easily found if the
data are adequate, which may not be the case for the three-dimensional
distribution of galaxies.  Some hints of the next term in the expansion
for small $r$ --- the lacunarity function (Thieberger {\it et al.}, 1990)
--- have been found for the galaxy distribution (Provenzale {\it et al.},
1997), though it has yet to be confirmed that the observed fluctuations
about the linear plot are intrinsic to the geometrical distribution of
galaxies.

\bigskip
\centerline{\bf 3. Projecting Fractals}
\medskip
\noindent

Having dealt with the polemical and pedagogical aspects of this
problem, we may now turn to our basic question: how are the
fractal properties of a distribution of points in
three-dimensional space transformed by projection of the set
onto a subspace like the celestial sphere? The main features
of this question can already be seen in projecting from two
dimensions into one dimension and, as this version of the problem
is more economical of computing resources, we adopt it here for
our illustrative examples.  The work for the full case of projection from
three dimensions is very much the same, as we have verified by performing
a few explicit tests.  By way of introduction to these problems the
reader may wish to look at the analytic study of a specific case,
performed by Mandelbrot (1975; see also Dogterom and Pietronero, 1991).

Before describing results obtained with actual examples, we recall
some of the relevant mathematical theorems.  As in most sciences,
we must then face the uncomfortable gap that arises between theory
and practice because the quantities most easily measured, here the $D_q$,
for $q>1$, are not those that conform best to mathematical usage.  For the
simplest systems, that is for the monofractals for which $D_q$ is
independent of $q$, these distinctions do not arise, while for
most practical cases they are slight.  So we shall not enter here
into the details; these are discussed in Beck and Schloegl (1993),
for example.

An empirically useful characterization of a set of points is by way of
its fractal dimension $D$ evaluated in a space of dimension ${\cal D}$.
To define this, we first find the smallest number of balls --- the
mathematician's word for the interiors of spheres of dimension ${\cal D}$
--- of diameter $\ell$ that are needed to cover all the points.  As $\ell
\rightarrow 0$, this number will vary like $\ell^{-D}$.  The
quantity $D$ is the fractal dimension of the set of points; it is
also called the box-counting dimension and it is almost always the
same as $D_0$. Computing it from data is less reliable than
determining $D_2$ but, if we know the recipe by which a
mathematically conceived set is constructed, it is not difficult
to find the exact values of $D_0$ and to compare them with those
determined from the exponents found by the use of correlation
functions (Thieberger, {\it et al.}. 1990).

For subtle reasons, for some sets that we will rarely encounter in
real life, the box-counting procedure just described may not work
and so the definition may be relaxed to allow balls of all sizes
in a covering of the set by balls. For each such generalized
covering, an effective dimension may be obtained and the least
upper bound of these is called the Hausdorff dimension, $D_H$. It
may be shown that $D_H \le D_0$.  The Hausdorff dimension is the
one that many mathematicians work with and, of course,
they also prefer to be more precise in defining it. However, we
almost never need to distinguish $D_H$ from $D_0$ (Mainieri, 1993),
which has an intuitive appeal and brings us closer to the way that de
Vaucouleurs (1970) thought about these things than to the more
conventional statistical treatments of galaxy distributions. (When asked
how he happened to employ such a seemingly mathematical approach, de
Vaucouleurs (private communication) explained that this was simply what
he had observed in the sky.) Although many of the theorems about fractal
sets are stated in terms of the Hausdorff dimension, Sauer and Yorke
(1995) showed that the theorems which apply to Hausdorff dimension apply
also to the correlation dimension, which may be shown to be a lower bound
on the Hausdorff dimension.

Since the actual examples we study here concern solely projections from a
two-dimensional point set onto a one-dimensional space such as a line or a
circle, we denote the original set by S2 and the projected set by S1.
Then we inquire into what can be said about the relation between these two
sets.

Generally speaking, one finds that $D_h (S1) = 1$, if $D_h (S2) >1$,
and that $D_h (S1) = D_h(S2)$, if $D_h (S2)<1$, as might have been
expected intuitively.  The theorems on which this rough summary is based
were already proven by Besicovitch (1939).  Moreover, Radons (1993) has
shown that, if $D_{\infty } (S2) <1$, not all $D_q (S1)$ are the same for
different $q$.  Hunt {\it et al.} (1997) proved a theorem whose relevant
part for us is that, under almost all linear transformations, from two
dimensions to one dimension, $D_2 (S1)= \min[1,D_2 (S2)]$.  In their
paper, they give a nonpathological example showing that for $q>2$ an
analogous equality does not hold.

Though the results just outlined as representing the theory of the
projection of fractals suggest how well-developed that theory is, we
must stress that it typically applies only when very large numbers of
points are involved.  Since it is sometimes true that the results of
asymptotic treatments do not accurately reflect the outcome of a real
situation with a finite number of points in the set, we have carried
out some explicit examples of the effects of projection on point sets.
We now describe these.  In choosing such examples, we must be aware that
projection onto a circle and onto a plane are fundamentally different.
When we project onto a plane here, we do it unidirectionally,
while our projections onto the unit circle will be along radii extending
from the center.  These differences demand special care in the numerical
work.  The sectors used in the projections onto the circle must be such
that a high enough density of points is encountered in each if we
are to obtain reproducible results.  And, in comparing different cases,
one should take care to use the same angles of projection.

To check on our procedures, we have verified that in projecting
homogeneous two-dimensional sets onto lines we obtain the
theoretical results to good accuracy, even for a sample of
moderate size. For fractals, however, the situation is more
complicated and to standardize the procedures we have worked with
Cantor fractals constructed as follows. We first make a Cantor set
by decimation.  That is, from the unit interval $[0,1]$ we remove
the open set $({1\over a}, 1-{1\over a})$, where $a>1$.  This leaves the
two outer segments $[0,{1\over a}]$ and
$[1-{1\over a},1]$.  The choice of the parameter $a$, which is $3$ for the
standard Cantor set, is left open for the moment.  A fractal set is then
constructed by repeating the procedure on the two remaining segments and
continuing in this way with each of the remaining segments in successive
generations. The result is a monofractal whose $D_q$ is $\log 2/\log
a$ for all $q$.

To construct a two-dimensional fractal we formed the Cartesian
product of two one-dimensional fractals, each with its own value of $a$.
We designate the set constructed in this way as $S2=(a_1,a_2)$. Such
composite sets are also monofractals whose dimension is the sum of
those of the two Cantorial sets (Falconer,1990):
 $$D(S2) = {\ln 2\over  \ln a_1} + {\ln 2 \over \ln a_2 } \ . \eqno(3.1)
$$
This fractal typically produces a different fractal dimension according
to the line on which it is projected.

Our calculations on these sets were performed using a procedure called
iterated function schemes (Barnsley, 1988).  Results for the projection of
three such multiplicative fractal sets are presented in the following two
tables. In Table 1, we give the projected dimensions for the three sets,
designated $(a_1,a_2)$, chosen to illustrate the projection effects for
cases with small, medium and large values of D(S2). In this table we
also show that projections onto a circle or onto a (generic) line give, to
good approximation, the same results, though care was taken to chose the
appropriate angles.

For the smallest number of points used in our analysis, we performed the
calculations with many different angles to limit our sample. Because of
the nature of projecting onto a circle, we obtained different projected
dimensions for different sets of angles. We then selected that set which
gave us the largest dimension. By using this procedure we eliminate
problematic regions (i.e. sparse regions) in our projection. We then
adopted this set of angles for the remainder of our calculations.

In Table 2, we show the convergence of the dimensions with
increasing numbers of points in the set, for the case of $D_2$.  In
seeking convergence, we doubled the number of points when going
from one line to the next.  Our results indicate that for the
fractal $S2=(6.54,3.)$ we do not get the value of two for $D(S1)$
as suggested in the aforementioned work of Hunt {\it et
al.} (1997).  Of course, we are dealing with finite sets and, from
these, it is not clear that we can achieve convergence to the
value predicted by theory. We also give in Table 2 the cases of
projecting a line and projecting a homogeneous (randomly
distributed) two-dimensional set on a circle. It is interesting
that we cannot distinguish between the results of these two
projections.
\medskip

\hskip 1cm{\bf Table 1: The Projected Dimensions.}

\vbox{\offinterlineskip
\hrule width 9.12cm
\halign{&\vrule#&
 \strut\quad\hfil#\quad
&\vrule#&\quad\hfil#\quad
&\vrule#&\quad\hfil#\quad
&\vrule#&\quad\hfil#\quad
&\vrule#&\quad\hfil#\quad\cr
height 2pt&\omit&&\omit&&\omit
&&\omit&&\omit&&\omit&\cr
&Set\hfil&&$D_{th}$&&$D_2$&&$D_4$
&&$D_6$&&N&\cr
height 2pt&\omit&&\omit&&\omit
&&\omit&&\omit&&\omit&\cr
\noalign{\hrule}
height 2pt&\omit&&\omit&&\omit
&&\omit&&\omit&&\omit&\cr
&(6.54,3.)&&1.000&&0.89&&0.86&&0.84&&62000&\cr
&(6.54,3.)\rlap*&&1.000&&0.88&&0.83&&0.79&&62000&\cr
&(15.,14.)&&0.519&&0.52&&0.50&&0.50&&58000&\cr
&(3.,2.68)&&1.334&&0.98&&0.98&&0.97&&58000&\cr
height 2pt&\omit&&\omit&&\omit
&&\omit&&\omit&&\omit&\cr}
\hrule width 9.12cm  *This case is projection on a line.}
\par
\medskip

\centerline{\bf Table 2: Variation of the calculated $D_2$ with N.*}
\par

\vbox{\offinterlineskip
\hrule width 12.8cm
\halign{&\vrule#&
 \strut\quad\hfil#\quad
&\vrule#&\quad\hfil#\quad
&\vrule#&\quad\hfil#\quad
&\vrule#&\quad\hfil#\quad
&\vrule#&\quad\hfil#\quad\cr
height 2pt&\omit&&\omit&&\omit
&&\omit&&\omit&&\omit&\cr
&N\hfil&&(6.54,3.)&&(15.,14.)&&(3.,2.68
&&line&&2D random&\cr
height 2pt&\omit&&\omit&&\omit
&&\omit&&\omit&&\omit&\cr
\noalign{\hrule}
height 2pt&\omit&&\omit&&\omit
&&\omit&&\omit&&\omit&\cr
&15&&.876$\pm $.007&&.520$\pm $.007&&.969$\pm $.003
&&.972$\pm $.001&&.975$\pm $.002&\cr
&30&&.883$\pm $.003&&.519$\pm $.002&&.978$\pm $.002
&&.985$\pm $.001&&.987$\pm $.001&\cr
&60&&.885$\pm $.003&&.519$\pm $.002&&.984$\pm $.001
&&.993$\pm $.001&&.993$\pm $.001&\cr
&120&&.888$\pm $.001&&.518$\pm $.002&&.987$\pm $.001
&&.996$\pm $.001&&.996$\pm $.001&\cr
height 2pt&\omit&&\omit&&\omit
&&\omit&&\omit&&\omit&\cr}
\hrule width12.8cm *The number of points,N, is given in thousands}

\bigskip
\centerline{\bf 4. Luminosity effects.}
\medskip
\noindent

In considering the distribution on the celestial sphere of galaxies or of
radio sources the role of the intrinsic luminosities of the observed
objects must be allowed for (Maurogordato and Lachi\`eze-Rey, 1991).
Such effects appear in $N(>f)$, the number of galaxies detected whose
flux density exceeds a given value, $f$ (Peebles, 1993).
This function is a more reliable indicator of the dimension of the set
under study than the dimension obtained directly from
projection without regard to luminosity effects.

As is often done in first approximation in the stellar case, we
assume that the luminosity function of the objects in our
study is independent of position and of time.  Though such
dependences (especially the latter) may be important and ought to
be included in more advanced studies, they involve uncertainties
and, in any case, we prefer to take things one at a time as we
learn how the various influences make themselves felt.

Then, for a homogeneous distribution of galaxies with no
evolution, we have (Peebles, 1993) $$ N(>f)= K f^{-1.5} \ .
\eqno(4.1)  $$
The constant, $K$, in this relation incorporates
the luminosity function, which for galaxies, is often chosen to be
the function (Schechter, 1976), $$ \sigma \propto w^\alpha \exp{(-w)}
\eqno(4.2)  $$
where $w=L/L_*$, $L$ is the luminosity, $L_*$ is
some characteristic luminosity and obervations give
$\alpha = -1.07 \pm 0.05$.

Equation (4.1) is the consequence of integrating all the points
over the range from zero to infinity.  In that case, the value of
$\alpha $ influences only the constant and not the power of $f$ in
(4.1).  In the case where one can write down the fractal
distribution in the same way as the homogeneous one, the equation
becomes (Peebles, 1993) $$ N(>f)= k  f^{-\beta }  \ .   \eqno(4.3)
$$ where $\beta $ can be approximated by $D_2/2 $.  The derivation is
approximate and does not allow for lacunarity oscillations (Provenzale et
al., 1997) that are washed out in projection according to Durrer
{\it et al.} (1997).

In the calculations described here, we project a two dimensional space
onto a one-dimensional one.  We nevertheless keep the same luminosity
versus distance law as for the three dimensional case, namely
$$ f = {L\over 4 \pi r^2} \ .  \eqno(4.4)  $$
Therefore, instead of eq. (4.1), we obtain the form
$$ N(>f)= K  f^{-1} \ .  \eqno(4.5) $$

We also consider the case of projecting a line onto a circle. To explain
the results in this case, we have to go into more detail concerning
how one obtains the equations. According to Peebles (1993) we can write
$$ N(>f)=\int^{\infty }_0 r^2 dr \psi (A r^2f) \  \eqno(4.6) $$
where
$$ \psi (x)=n\int^{\infty }_x w^{\alpha } \exp^{-w} dw \  \eqno(4.7)   $$
and $\alpha $ is the exponent from Eq. (4.2).

In the case of the projection of a line, where the variation in r is not
very large (for example, for the case of Table 2: $84>r>71$), the limits
will depend on f.  Therefore we cannot assume that we can incorporate the
luminosity
function into K as in the other cases. Now K will depend on the
luminosity function, that is, on the value of $\alpha $.

To illustrate our point let us assume that we could approximate the
different r values by their means.  Then we could introduce
a $\delta$-function into the integral with argument $(r-r_m) $ so that
$$ N(>f)=\int^{\infty }_0 rdr \delta (r-r_m)\psi (Afr^2)=r_m
\psi (Afr_m^2) \ .  \eqno(4.8) $$
Thus we can expect the power of f to be influenced by $\alpha$, as
confirmed
by our numerical calculations.  It is therefore convenient to write
$$ N(>f)={\rm const.}\;  f^{\beta} \   \eqno(4.9)  $$
and to evaluate the parameters in this expression from the calculations.
\par
\medskip

\centerline{\bf Table 3: Dependence of $N(>f)$ on $f$.}
\par

\vbox{\offinterlineskip
\hrule
\halign{&\vrule#&
 \strut\quad\hfil#\quad
&\vrule#&\quad\hfil#\quad
&\vrule#&\quad\hfil#\quad
&\vrule#&\quad\hfil#\quad
&\vrule#&\quad\hfil#\quad\cr
height 2pt&\omit&&\omit&&\omit
&&\omit&&\omit&&\omit&\cr
&Set type\hfil&&Specification&&$D_{th}$&&$\beta_{th}$
&&$\beta_{cal}$&&$\alpha $&\cr
height 2pt&\omit&&\omit&&\omit
&&\omit&&\omit&&\omit&\cr
\noalign{\hrule}
height 2pt&\omit&&\omit&&\omit
&&\omit&&\omit&&\omit&\cr
&random&&2D&&2.00000&&-1.000&&-0.99$\pm $0.01&&-1.25&\cr
&random&&2D&&2.00000&&-1.000&&-1.00$\pm $0.01&&-3.25&\cr
&random&&1D&&1.00000&&---&&-2.0$\pm $0.1&&-1.25&\cr
&random&&1D&&1.00000&&---&&-2.7$\pm $0.1&&-3.25&\cr
&cantor&&(6.54,3.)&&1.00003&&-0.500&&-0.51$\pm $0.03&&-1.25&\cr
&cantor&&(6.54,3.)&&1.00003&&-0.500&&-0.51$\pm $0.02&&-3.25&\cr
&cantor&&(4.7,2.7)&&1.14575&&-0.573&&-0.57$\pm $0.02&&-1.25&\cr
&cantor&&(7.,4.)&&0.85621&&-0.428&&-0.41$\pm $0.05&&-1.25&\cr
height 2pt&\omit&&\omit&&\omit
&&\omit&&\omit&&\omit&\cr}
\hrule}

\medskip
In Table 3, we compare the $\beta $ obtained for the different
sets on which we performed our calculations.  The values are
obtained by least square fit of $N(>f)$ for different $f$ values.
As $f$ increases, the number of rejected points increases. The
higher the $r_{max}$ the fewer retained points remain out of a
given sample. The larger $f$, the steeper this decrease is. In the
calculations used for Table 3, we kept the number of projected
points at the constant value of 40000 throughout by appropriately
increasing the number of points processed as the number of
rejected points rose.  Even a smaller sample gives quite
reasonable results.

As our $R_{max}$ is perforce finite, we have
to take care not to include too small values of $f$. If this point
is ignored we obtain a more complicated dependence of $N(>f)$ on
$f$.  To get the appropriate range where $R_{max}$ does not
influence the results, we look for the $R$ for which almost no
projected points reach the unit circle, for a given $f$. Only
those $f$ that are not eliminated because of $R_{max}$ are then
included in the calculations.  We also examined the influence of
the results on $\alpha $ (Eq. (4.2)). As might be expected,
$\alpha $ does not influence the results except in the case of
projecting a line onto a circle. Our results are in good agreement
with the analysis described by Peebles (1993).

\vskip 20 pt
\centerline{{\bf 5.  Conclusion}}
\bigskip

We conclude that extensive two-dimensional catalogues of extraterrestrial
objects can reveal more about the statistics of these objects than may
have been supposed.  We have illustrated the possibilities with a simple
exercise in fractal projection that shows how the existing mathematical
results on this subject may be brought to bear on the study of large-scale
structure.  Though we have included only one feature not explicitly
mentioned in most mathematical studies, namely the luminosity
distribution, we see already the emergence of an interesting aspect of the
projection process.  To appreciate the usefulness of this result, we
should recall that, in the projection of a homogeneous (or randomly
distributed) set, there is no real difference in the dimension between the
original and the projected dimension once the change in embedding
dimension is allowed for.  However, if a suitable luminosity function is
folded into the projection process, pronounced differences emerge.  This
makes it possible to distinguish among sets with different dimensions
after projection through the influence of the luminosity function.

Thus even when plentiful three-dimensional data become available,
projection effects will be worth studying since they avoid some of the
problems in the measurement and interpretation of redshifts, while still
revealing the influence of the luminosity function.  For example, the
three regimes mentioned in Murante {\it et al.} (1998), give appreciably
different results on projection if the luminosity effects are
included.  It is also possible to include the effects of source
evolution or of galaxy merger by making the parameters in the Schechter
distribution time-dependent.

In conclusion, we thank Antonello Provenzale for his critical reading
of the manuscript.  We are also grateful to S. Bhavsar, I. Grenier
J. Yorke for for providing some references.  E.A.S. would like to express
his gratitude to the Tata Institute for Fundamental Research for
hospitality in March, 1999 with special thanks to Kumar Chitre for his
many kindnesses during that visit.

\bigskip
\centerline{\bf References}
\bigskip
{\eightpoint\parindent=0pt\everypar={\hangindent=0.5 cm}
\noindent Barnsley M.F., 1988, {\it Fractals Everywhere},
Addison-Wesley, Reading, MA.

\noindent Besicovitch A.S., 1939, Math. Annalen, 116, 349.

\noindent Becker C. and Schloegl F., 1993, {\it Thermodynamics of
chaotic systems}, Cambridge Univ. pp. 97-101.

\noindent Bessis D. and Fournier, J.-D., 1989, Phys. Lett. A, {\bf 140},
331.

\noindent Borgani S., 1995, Phys. Rep., 251, 1.

\noindent Coleman P.H., Pietronero L. and Sanders R.H., 1988,
Astron. \& Astrophys.,
200, L32.

\noindent Coleman P.H. and Pietronero L., 1992, Phys. Rep.,
231, 311.

\noindent Cress, C.~M., Helfand, D.~J., Becker, R.~H., Gregg, M.~D.
and White, R.~L., 1996, Ap. J. 473, 7.

\noindent Cress, C.~M. and Kamionkowski, M., 1998, MNRAS, ???.

\noindent Dogterom M. and Pietronero L., 1991, Physica A., 1991,
171, 239.

\noindent Durrer, R., Eckmann, J.-P., Labini, S.~L., Montouri, M.
and Pietronero, L., 1997, Europhys. Lett., 40, 491.

\noindent Falconer K., 1990, {\it Fractal Geometry}
(Chichester, UK: John Wiley and Sons)

\noindent Feder J., 1988, {\it Fractals}, Plenum Press, N.Y., U.S.A.

\noindent Grassberger P. and Procaccia I., 1983, Phys. Rev. Lett.,
50, 346.

\noindent Groth E.J. and Peebles P.J.E., 1977, Astrophys. J., {\bf 217},
385.

\noindent Halsey T.C., Jensen M.H., Kadanoff L.P., Procaccia
I. and Shraiman B.I., 1986, Phys. Rev. A, 33, 1141.

\noindent Hunt B.R. and Kaloshin V.Yu., 1997, Nonlinearity, 10, 1031.

\noindent Kerscher M., 1999, A\&A , 343, 333.

\noindent Mainieri, R., 1993, Chaos, 3, 119.

\noindent Mandelbrot B.B., 1975,  Comptes Rendus 280A, 1551.

\noindent Mandelbrot B.B., 1982, {\it The Fractal Geometry of Nature}
(San Francisco: Freeman).

\noindent Martinez V.J. and Jones B.J.T., 1990, Mon. Not. R.
Astron. Soc., 242, 517.

\noindent Maurogordato, S. and Lachi\`eze-Rey, M., 1991, Ap. J., 369, 30.

\noindent Murante G., Provenzale A., Borgani S., Campos A.
and Yepes G., 1996, Astroparticle Phys., 5, 53.

\noindent Murante G., Provenzale A. Spiegel E.A. and Thieberger R.,
1997, Mon.  Not.  R.  Astron.  Soc., 291, 585.

\noindent Murante G., Provenzale A. Spiegel E.A. and Thieberger R.,1998,
 Annals of the New York Academy of Sciences, 867, 258.

\noindent Neyman J., Scott, E.L. and Shane, C.D., 1953, Astrophys. J.,
{\bf 117}, 92.

\noindent Neyman J. and Scott E.L. 1959, Hdbch d. Phys., {\bf 53},
Berlin: Springer, p. 416, ff.

\noindent Paladin G. and Vulpiani A., 1984, Nuovo Cimento Lett., 41,
82.

\noindent Paladin G. and Vulpiani A., 1987, Phys. Rep., 156, 147.

\noindent Peebles P.J.E., 1980, {\it The Large Scale Structure of the
Universe} (Princeton: Princeton Univ. Press).

\noindent Peebles P.J.E., 1993, {\it Principles of Physical Cosmology}
 (Princeton: Princeton University Press) p. 215.

\noindent Provenzale A., 1991, in {\it Applying Fractals in Astronomy},
A. Heck and J.M. Perdang Eds. (Berlin: Springer).

\noindent Provenzale A., Spiegel E.A. and Thieberger R., 1997, Chaos,
7, 82.

\noindent Radons G., 1993, J. of Stat. Phys., 72, 227.

\noindent Renyi A., 1970, {\it Probability Theory} (Amsterdam:
North Holland).

\noindent Ribeiro, M.~B, 1995, Ap. J., {\bf 441} 477.

Ryle, M. and Clarke, R.~W., 1961, MNRAS, {\bf 122}, 329.

\noindent Sauer T.D. and Yorke J.A., 1995(?), in {\it Ergodic Theory and
Dynamical Systems}, ???

\noindent Schechter, P., 1976, Ap. J., {\bf 203}, 297.

\noindent Spedalere, R. and Schucking, E.~L., 1980, Astron. J.,
{\bf 85}, 586.

\noindent Thieberger R., Spiegel E.A. and Smith L.A., 1990, in
{\it The Ubiquity of Chaos}, S. Krasner Ed. (AAAS Press).

\noindent Trumpler, P.~J. and Weaver, H.~F., 1953 {\it Statistical
Astronomy} (Berkeley and Los Angeles: University of Calif.)

\noindent Zeldovitch Ya.B.,1970, Astrofizika(A) {\bf 6}, 319.

}                                         
\end